\documentclass[twocolumn,aps, prd]{revtex4}%
\usepackage{graphicx}
\usepackage{dcolumn}
\usepackage{bm}
\usepackage{amsmath}%
\usepackage{amsfonts}%
\usepackage{amssymb}
\begin{document}

\title{Interplanetary coronal mass ejection effect on the muon flux at sea level }

%\title{Solar wind effect on the muon flux at sea level}

\author{C. E. Navia}
\address{Instituto de F\'{\i}%
sica Universidade Federal Fluminense, 24210-130,
Niter\'{o}i, RJ, Brazil} 

\author{C. R. A. Augusto}
\address{Instituto de F\'{\i}%
sica Universidade Federal Fluminense, 24210-130,
Niter\'{o}i, RJ, Brazil}

\author{M. B. Robba}
\address{Instituto de F\'{\i}%
sica Universidade Federal Fluminense, 24210-130,
Niter\'{o}i, RJ, Brazil}

\date{\today}

\begin{abstract}
We present the results of $720$ hours of observations of transient solar events at ground 
level, during the summer season 2005 (Souther Hemisphere). 
Data were taken with the TUPI muon telescope, working at a high counting rate 
(up to $100$ KHz) and always pointing on the IMF lines (45 degrees of pitch angle). An 
anti-correlation among the arrival of keV protons (observed by EPAM detector aboard the ACE spacecraft) 
and sudden depressions in the muon flux at sea level have been observed. The phenomena 
is discussed in the context that they can be considered as mini-Forbush, caused by a 
shielding effect of the passage of a disturbance (shock and plasma) and may be a signature of 
interplanetary  manifestations of coronal mass ejections.
\end{abstract}


\maketitle

Direct measurements of solar energetic particles have been made successfully 
using satellite-borne observatories. However, these measurements are 
limits to the MeV energy region by the small active areas in space. 
The high energy solar particles in the MeV to GeV energy region or above can be 
obtained using only indirect methods such as ground-based detectors. The ground-based 
detectors can infer information about the primary solar particles only from the showers 
originating from their interaction with air nuclei. This makes such 
observations extremely dependent on the knowledge of the shower 
development in the atmosphere. The detection of solar neutrons and charged particles 
at ground level using the neutron monitor world-wide network starting from 1954 by Simpson
\cite{simpson54} have shown an excellent performance, 
because the intensities are recorder to several geomagnetic cut-offs and the 
geomagnetic dependence, anisotropies and other characteristics may be better known.
Transient depressions reaching a minimum value in approximately one day in the galactic cosmic 
ray intensity  following for a gradual recovery in up to several days have been observed in 
neutron monitors data. These phenomena, called "Forbush events" \cite{forbush37}
have been known over the past 6 decades. 
It is now known that they are associated with the passage of an interplanetary disturbance 
(shock and plasma enveloping the Earth) \cite{parker63} and in most cases they are considered 
as an interplanetary manifestations of coronal mass ejections.

On the other hand, in the framework of a study of a possible "photo-muon" excess from galactic 
center, several ground level enhancements (GLEs) in the muon flux have been found. Our previous 
paper was devoted to cross-correlation analysis between the sudden commencements in the 
muon counting rate at sea level and the X-ray prompt emission of solar flares of small scale
\cite{augusto2005}. 
The high performance of the TUPI muon telescope with respect to a muon cluster whose origin is the arrival
in the upper atmosphere of a small bundle of protons and/or ions with energies exceeding 
the pion production and above the local geomagnetic cut-off ($9.8$ GV) arises mainly from:
(1) its high counting rate ($\sim 100$ kHz) and (2) its tracking system. The telescope is 
always looking near to the direction of the IMF lines. 
Details of the experimental setup of the TUPI telescope have been reported in  
\cite{augusto2003,navia2005}.

We present here the results of a experiment dedicated to the investigation of 
transient solar events such as solar flares and their influence on the galactic cosmic 
rays at 1 AU. The $720$ hours of observation during the summer 
season 2005 (Souther Hemisphere) correspond to $60$ raster scans
with a duration of 12 hours each, across parallel lines in declination (Sun's declination)
and an hour angle of 3 hours early in relation to the Sun's hour
angle, under this conditions
the pitch angle is always $45^0$ (where $0^0$ pitch angle represent sun-ward direction).
Figure 2 from the ref. \cite{augusto2005}. 
summarizes the situation.
In this summer session we have identified  several sudden depressions (already observed previously)
in the muon counting rate and we show here that these muon intensity depressions
and the arrival of solar protons in the keV energy region at the ACE spacecraft \cite{ace}
may be a good indicator of the passage of a small interplanetary disturbance (shock and plasma).
This is because sudden and/or gradual increases of keV protons (above the proton flux background flux)
in the ACE spacecraft coincide with the beginning of the depressions in the TUPI muon intensity at
ground level.

On the other hand, ref. \cite{cane96} has reported Forbush events in 30 years of neutron 
monitor data. Forbush events was observed also in spacecraft experiments \cite{cane97},  
in air shower arrays \cite{inoue01}, as well as in muon telescopes at ground level
\cite{fujimoto01}, the magnitudes of the Forbush depression depend of several factors such as 
the energy threshold of the detected particles, the 
geomagnetic cut-off of the observation location, 
as well as how the data are presented. For instance hourly averages present bigger magnitudes 
than the daily averages. In the TUPI experiment the Forbush events have been observed as a drastic 
change in the muon count rate (raw data). The muons detected ($E_{\mu}>0.1$ GeV) are produced by 
primary charged particles (protons and/or ions) with an energy above the geomagnetic cut-off (9.8 GV) 
and the data is presented in bins of 15 minutes. 
Figures from 1 to 5 presents five samples of events observed during the summer   
season 2005. In all cases, in the lower panels,  the TUPI muon relative intensity   
(pressure corrected) is presented, and in the upper panels   
the solar proton flux (EPAM protons) as observed by ACE spacecraft in the keV energy   
band is shown. From these figures it is possible to extract some especial characteristics of Forbush 
events observed by a narrow angle ($9.5^0$ of aperture) muon telescope at sea level, with a 
geomagnetic cut-off of 9.8 GV and always pointing to the IMF lines (45 degrees of picht angle).
These can be summarized by the following:
(a) In most cases the sudden depression coincide with the arrival at the Earth of 
solar protons (EPAM protons)in the keV energy band. In the event on 2005/02/24 the arrival of keV 
protons is around 40 minutes before the depression and in the event on 2005/04/26 the arrival
of keV protons is $\sim 2$ hours before the depression.
(b)In all cases the duration of the depression 
exceed the 23 hours UT when the telescope tracking run is off, and we do not have the duration 
of the depressions, we have only some evidence 
indicating a recovery time shorter than 24 hours. However, in order to obtain this information 
we include in the analysis two events observed out of the summer season 2005, in two 
consecutive days. Figure 6 summarize the situation. All the characteristics (as above mentioned) 
of Forbush events observed in the TUPI muon flux are valid for these two events. In addition the recovery 
time can be obtained as  21 hours. This short recovery time contrast with the recovery time 
(several days to week) observed in Forbush events in NM data.
(c)In some cases before the sudden depression a pre-Forbush
increase in the muon flux is observed, probably caused by the galactic cosmic ray 
acceleration at the front of the advancing disturbance. 
A clear pre-Forbush increases can be observed for instance in the event
on 2005/03/03 (see fig.4) and specially on 2005/04/25 (see Fig.6 right). A big pre-Forbush as shown 
in fig.6 may be  also a signature  
of the presence of GV protons (above the geomagnetic cut-off) associated with the interplanetary 
disturbance \cite{humble95}. However this type of signature (GV associated protons) can be observed 
better in the polar regions where the geomagnetic cut-off is smaller.
 (d)In contrast with Forbush decrease events as observed in neutron monitor (NM) data, 
characterized with a sudden depression lasting 10 hours to one day to reach a minimum 
intensity, the Forbush events observed in the TUPI muon flux have a sudden remarkable 
depression reaching a minimum intensity around one hour. 
(e)In the case of the event on 2005/03/15 (Fig.5) there is also the arrival the very high 
energy proton flux in the MeV energy band
(ACE SIS-protons) around 1.5  hours before the beginning of the depression. The 
prompt high energy solar particles observed inside or immediately before of keV solar proton 
(ejecta) suggest that the ejecta field lines are connected to the Sun.
(f)The event on 2005/04/25 (see Fig.6) is possible to see the two step signature of the 
Forbush depression, the first decrease occurs due to the passage of the turbulent field (shock) 
followed by the passage of the ejecta (plasma). It is possible to see also which the magnetic cloud
is not uniform during its passage to the earth, because anisotropies can be observed in the 
depressions (see figures 5 and 6 right).
  
On the other hand, if Forbush were the interplanetary manifestation of CMEs  
a $1:1$ correspondence will be expected.
However, if we take into account only those cloud (shock and ejecta) crossing the 
vicinity on Earth a correspondence between CMEs and Forbush of $7:1$ would be expected
\cite{cane2000}. The occurrence
rates of CMEs at solar minimum is $\sim 0.7$ per day, meaning that the occurrence rates of 
Forbush on Earth is $0.1$ per day or $3$ per month. A preliminary result from the present 
experiment it indicates a Forbush rates of $9$ per month, three times larger than the expected 
value. This results suggest a second cause or mechanism for the origin of the Forbush events, such as
co-rotating high speed streams \cite{iucci79},
because, the solar wind can be followed by a fast moving stream, the faster moving material  
will catch-up to the slower material and this interaction produces shock waves that can accelerate 
particles. One might question also the ability of ground-based experiments 
to detect all Forbush events, because only a fraction of Forbush (22 events) 
\cite{penna2005} were identified in
57 ICME-CME pairs \cite{gopalswamy2001}. However, any conclusion with respect to Forbush rates is 
premature. Our result need further confirmation because the present 
ongoing experiment will deliver much better statistics in the coming years.

Besides the high rate of Forbush events observed in the TUPI experiment, the change in the 
temporal scale when compared with NM events is other characteristic observed in TUPI events. 
The beginning of the TUPI Forbush events correspond, in a straightforward way, to local minima   
in the hourly average flux observed by the Moscow Neutron Monitor  \cite{moscow}. However, during the TUPI 
Forbush events the $k_p$ scale \cite{noaa} is smaller than 5. The $k_p$
index of 0 to 4 is below magnetic storm, consequently there is not storm notification (or storm level G0) 
at least according to the NOAA Space Weather Scale.
Consequently, we are probably in the front of a new category of Forbush events, 
and we called mini-Forbush events.
They happen in the high energy region and in
anti-coincidence with the arrival on the Earth of keV protons 
and this characteristic can be used as a signature of an interplanetary "thin" disturbance crossing the 
vicinity of Earth.

This work was partially supported by FAPERJ (Research foundation of the State
of Rio de Janeiro) in Brazil. The authors wish to express their thanks to Dr.
A. Ohsawa from Tokyo University for help in the first stage of the
experiment and to Dr. M. Olsen for reading the manuscript. 
We are also grateful to the various catalogs available  on the internet and to 
their open data police, especially to the ACE Real-Time Solar Wind (RTSW) Data.

%%%%%%%%%%%%%%%%%%%%%%%%%%%%%%%%%%%%%%%%%%%%%%%%%%%%%%%%%%%%%%%%%%%%%%%%%%%%%%%%%%%%%%%%
%%%%%%%%%%%%%%%%%%%%%%%%%%%%%%%%%%%%%%%%%%%%%%%%%%%%%%%%%%%%%%%%%%%%%%%%%%%%%%%%%%%%%

%%%%%%%%%%%%%%%%%%%%%%%%%%%%%%%%%%%%%%%%%%%%%%%%%%%%%%%%%%%%%%%%%%%%%%%%%%%%%%%%%%%%%%%% 

\begin{figure}[th]
\includegraphics[clip,width=0.6
\textwidth,height=0.6\textheight,angle=0.] {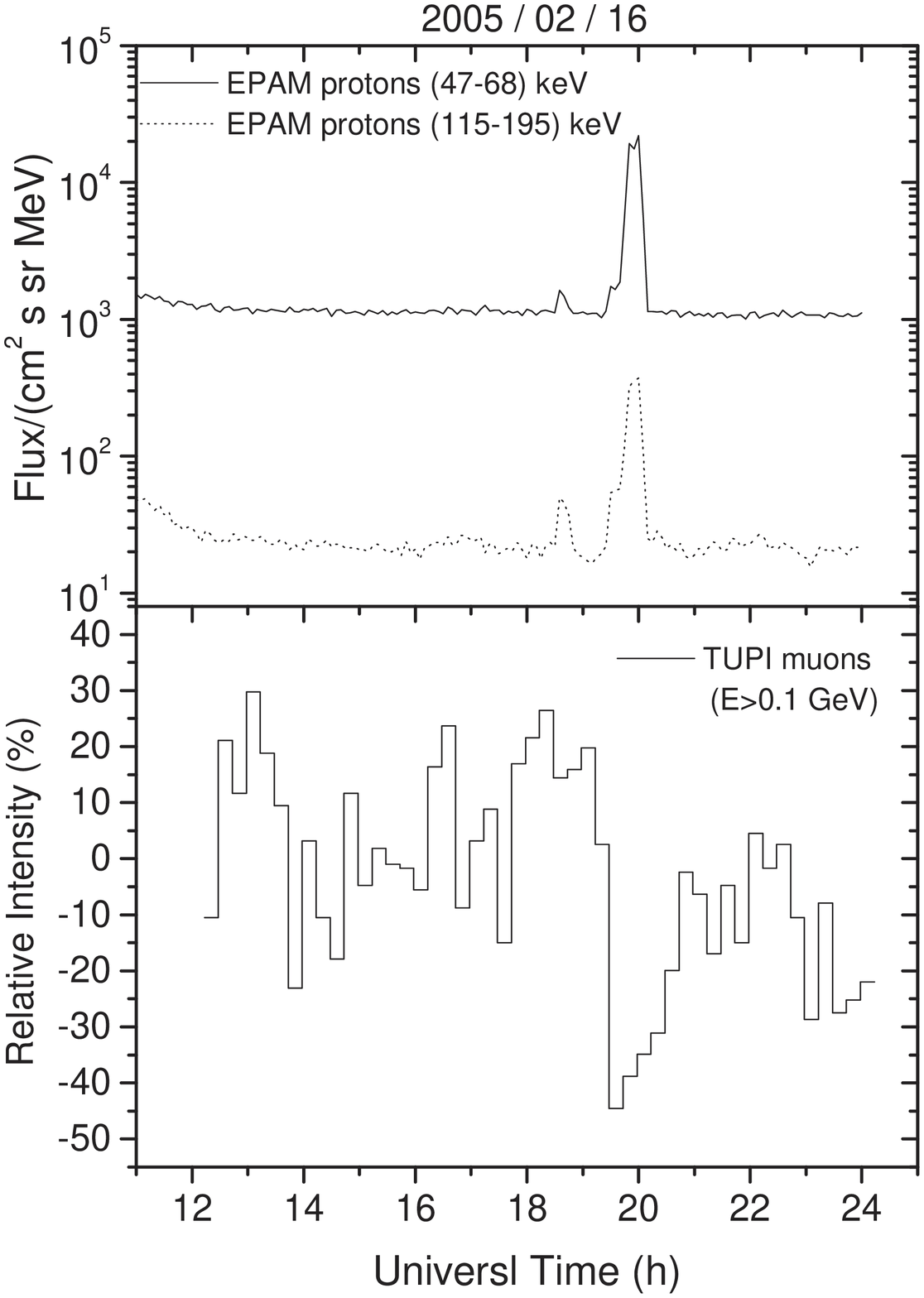}
\caption{Upper panel: Time profiles of the EPAM protons in two different energy bands.
Lower panel: TUPI muon relative intensity after pressure correction for the 2005/02/16 raster scan.}%
\end{figure}

\begin{figure}[th]
\includegraphics[clip,width=0.6
\textwidth,height=0.6\textheight,angle=0.] {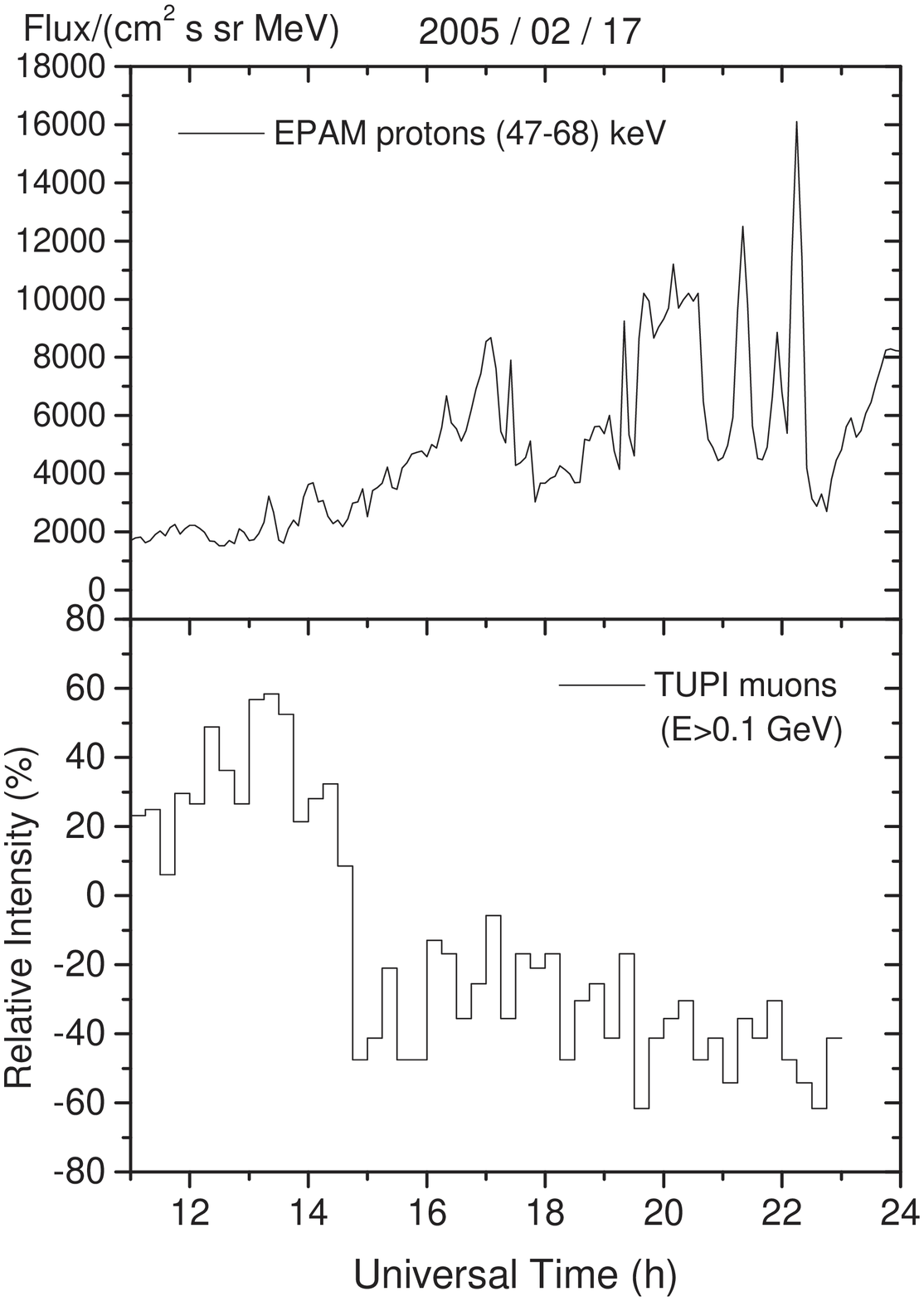}
\caption{The same as  figure 2, only for the raster scan on
2005/02/17.}%
\end{figure}

\begin{figure}[th]
\includegraphics[clip,width=0.6
\textwidth,height=0.6\textheight,angle=0.] {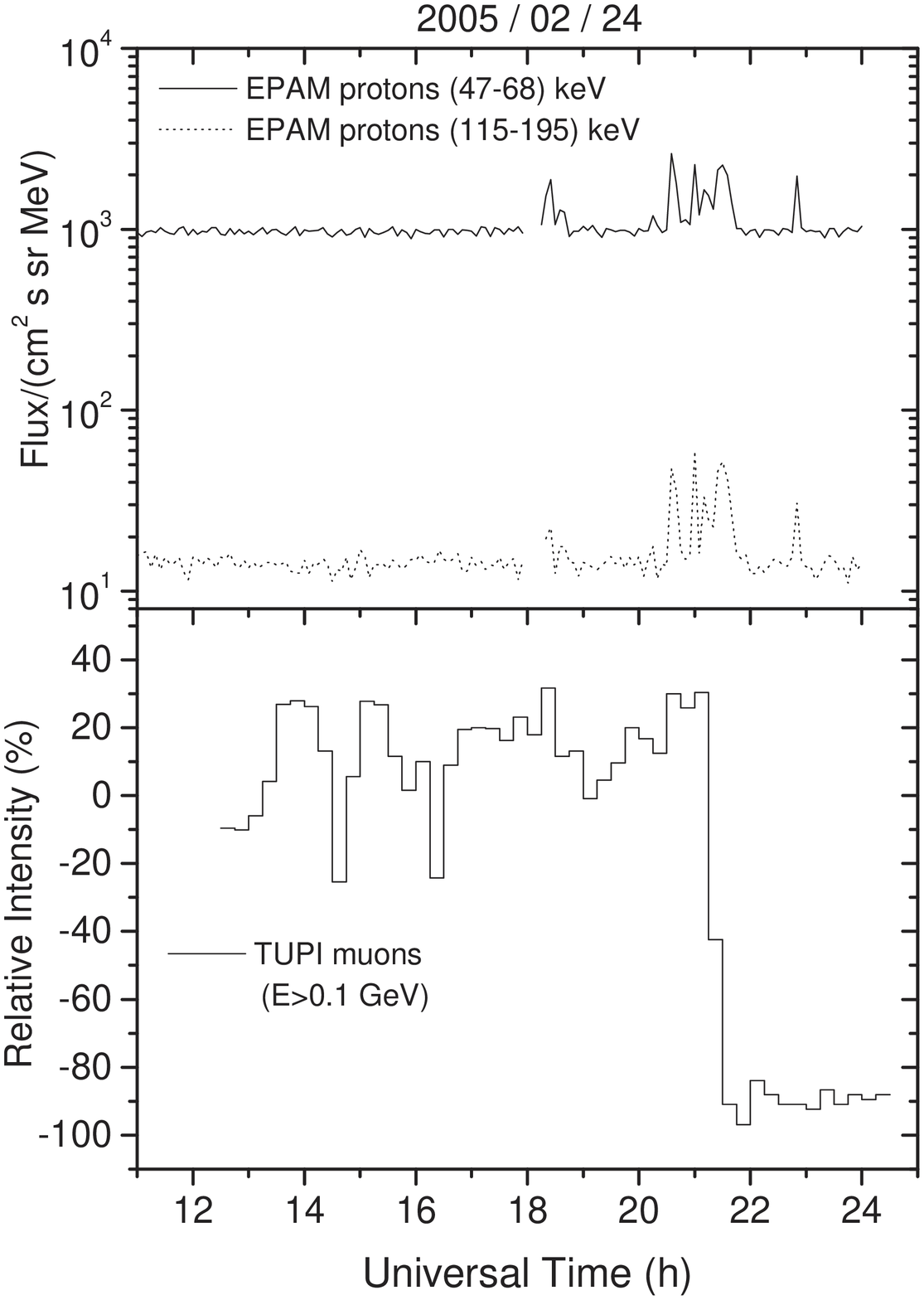}
\caption{The same as  figure 2, only for the raster scan on
2005/02/24.}%
\end{figure}

\begin{figure}[th]
\includegraphics[clip,width=0.6
\textwidth,height=0.6\textheight,angle=0.] {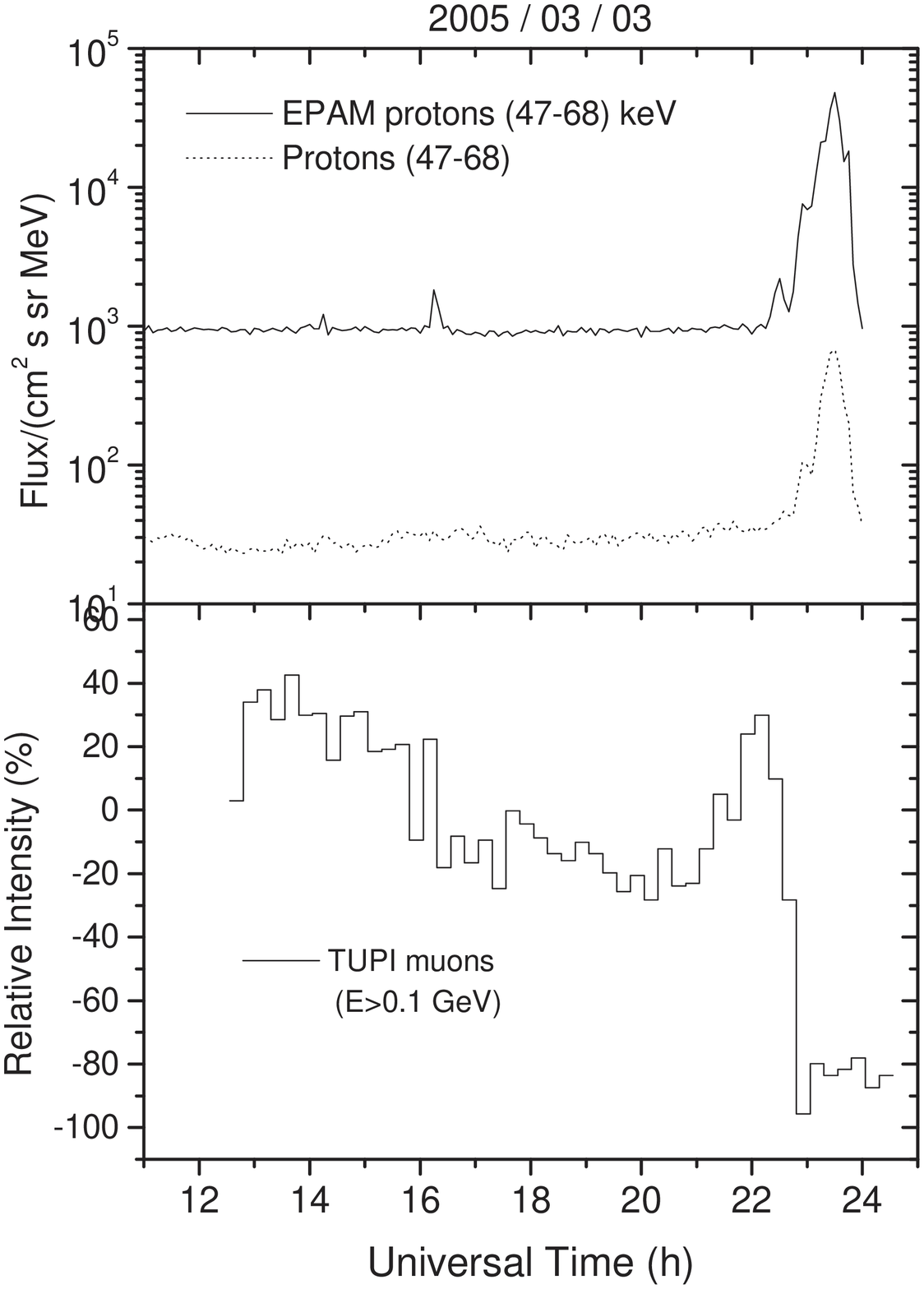}
\caption{The same as  figure 2, only for the raster scan on
2005/03/03.}%
\end{figure}

\begin{figure}[th]
\includegraphics[clip,width=0.6
\textwidth,height=0.6\textheight,angle=0.] {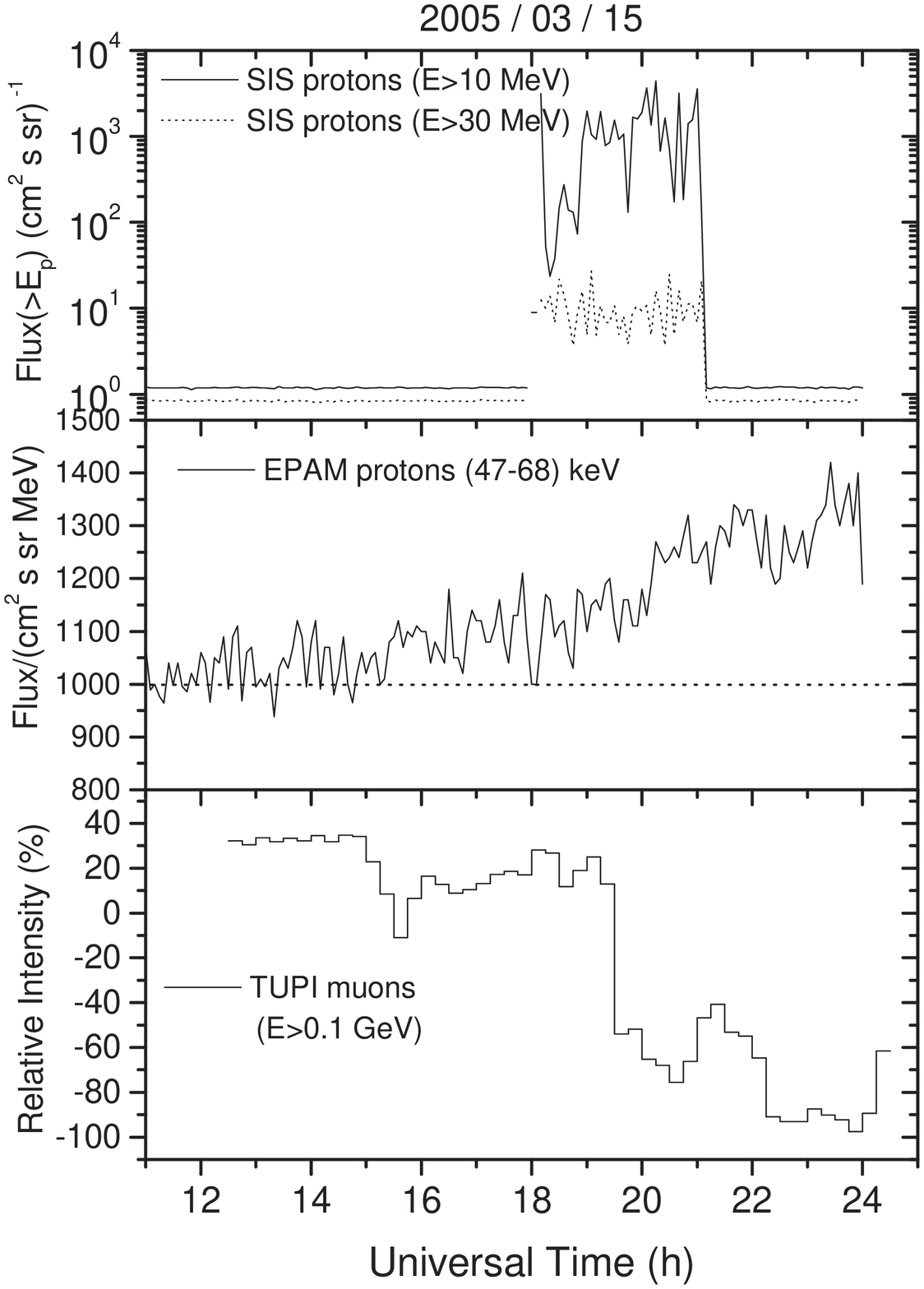}
\caption{The same as  figure 2, only for the raster scan on
2005/03/15.}%
\end{figure}

\begin{figure}[th]
\includegraphics[clip,width=0.6
\textwidth,height=0.6\textheight,angle=0.] {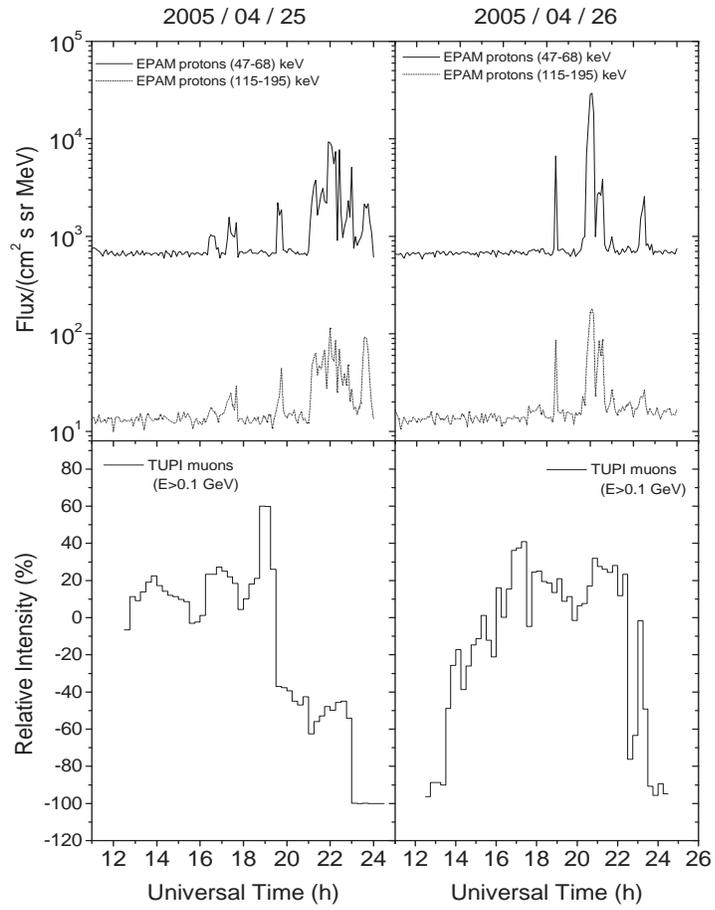}
\caption{The same as  figure 2, only for the raster scan on
2005/04/25 and 2005/04/26.}%
\end{figure}

\end{document}